\documentclass[12pt]{article}

\usepackage{amsfonts}
\usepackage{amssymb}
\usepackage{amsmath}

\usepackage{graphicx}
\usepackage{latexsym}
\usepackage{color}
\usepackage[colorlinks=true]{hyperref}
\usepackage{authblk}

\usepackage{graphicx,tikz}
\usepackage{amsmath,amssymb,mathrsfs}

\def\be{\begin{equation}}
\def\ee{\end{equation}}
\def\bea{\begin{eqnarray}}
\def\eea{\end{eqnarray}}
\def\bean{\begin{eqnarray*}}
\def\eean{\end{eqnarray*}}

\def\scri{\mathscr{J}}

\newlength{\cellwidth}
\begin{document}
\hfill YITP-22-104
\title{Ultra-massive spacetimes}

\author{Jos\'e M. M. Senovilla$^{1,2,3}$}
\affil{$^1$Departamento de F\'isica, Universidad del Pa\'is Vasco UPV/EHU, Apartado 644, 48080 Bilbao, Spain\\
$^2$Yukawa Institute for Theoretical Physics,
Kyoto University, 606-8502, Kyoto, Japan.\\
$^3$ EHU Quantum Center, Universidad del Pa\'{\i}s Vasco UPV/EHU.}

{\let\newpage\relax\maketitle}

\vspace{-0.2em}

\begin{abstract}
A positive cosmological constant $\Lambda >0$ sets an upper limit for the area of marginally future-trapped surfaces enclosing a black hole (BH). Does this mean that the mass of the BH cannot increase beyond the corresponding limit? I analyze some simple spherically symmetric models where regions within a dynamical horizon keep gaining mass-energy so that eventually the $\Lambda$ limit is surpassed. This shows that the black hole proper transmutes into a collapsing universe,  and no observers will ever reach infinity, which dematerializes together with the event horizon and the `cosmological horizon'. The region containing the dynamical horizon cannot be causally influenced by the vast majority of the spacetime, its past being just a finite portion of the total, spatially infinite, spacetime. Thereby, a new type of horizon arises, but now relative to past null infinity: the boundary of the past of all marginally trapped spheres, which contains in particular one with the maximum area $4\pi/\Lambda$. The singularity is universal and extends mostly outside the collapsing matter. The resulting spacetimes models turn out to be inextendible and  globally hyperbolic. It is remarkable that they cannot exist if $\Lambda$ vanishes. Given the accepted value of $\Lambda$ deduced from cosmological observations, such ultra-massive objects will need to contain a substantial portion of the total {\it present} mass of the {\it observable} Universe.
\end{abstract}

\section{Introduction}
Since the beginning of the century we know that the observable Universe is in accelerated expansion which implies the existence of a {\em positive} cosmological constant $\Lambda >0$ \cite{Riess1998,Perlmutter1999}. It is also known that a positive $\Lambda$ imposes restrictions on the area of marginally (outer) future-trapped surfaces \cite{HSN,W} if these are spatially stable in the sense of \cite{AMS,AMS1} --equivalently, `outer' in the sense of \cite{Hay,HSN}-- if the dominant energy condition holds. These limits can be generalized and strengthened by adding electromagnetic charge \cite{Simon}. The stability assumption can be understood as stating that the marginal trapped surfaces (MTS) enclose a black hole (BH) region. The area $A$ of (spatially stable) MTS is limited by
\be\label{lim}
A< \frac{4\pi}{\Lambda} .
\ee

Taking into account the relationship between the area of MTS (and of black hole event horizons) and their mass, one may wonder which kind of mechanism, if any, prohibits a BH with $A$ near the limit \eqref{lim} to increase its area by simply receiving more mass-energy from its exterior. Observe that this is completely different from the known cases of over-spinning or over-charging BHs with the goal of producing naked singularities, since in those cases there exist repulsive forces and a struggle between the increase in charge and/or angular momentum and the associated increase of the mass \cite{SMilestone}: the theory seems to conspire so that cosmic censorship prevails \cite{Wald1}. In contrast, in principle a BH will simply become bigger by adding mass, and it is difficult to imagine what can prevent such physical process.

In this paper, in order to understand this problem, I consider some simple models of spherical BHs that keep increasing their masses until the stable MTS of spherical topology reach the area-limit value and beyond. I will analyze the simplest possible models, first based on the Vaidya-de Sitter metric \cite{VS,Mallet} and then also in combination with the $\Lambda >0$ generalizations of the Oppenheimer-Snyder and Lema\^\i tre-Tolman collapses studied in \cite{GV,N,MS,L,DJCJ} --later re-discovered in \cite{SA,SA1}. In both cases we show that there is no problem in having larger and larger masses, but the dynamical horizon foliated by marginally trapped spheres then simply ends its existence. The cosmological horizon totally vanishes. The global structure of the resulting spacetimes is shown in convenient conformal diagrams. 

The global nature of event horizons is partly behind its dematerialization in these extreme spacetimes that I have called {\em ultra-massive}. However, that is not the main reason, or at least not only: the vanishing of future null infinity $\scri^+$ is the basic fact. This absence leads to `frustrated event horizons' and, as was to be expected, the area limit  \eqref{lim} is never surpassed by any MTS that is stable in spacelike directions. 

The next section is devoted to understanding the basic properties of Vaidya-de Sitter spacetimes. They can be easily inferred from those of the Kottler metrics (also known as Schwarzschild-de Sitter), which are well known and thoroughly studied in the literature. However, for the benefit of the reader I have added a useful Appendix with the main properties and corresponding conformal diagrams of the Kottler metrics. They can help in better understanding the main text. In section \ref{sec:first} I present the first type of models, based on Vaidya-de Sitter exclusively. Section \ref{sec:second} is devoted to the second type of models, which combine the first type of models with black holes in formation by stellar collapse using the Oppenheimer-Snyder-de Sitter models and others. I end the paper with an extensive discussion.

\section{The Vaidya-de Sitter metric}
Using the advanced null coordinate $v$ the Vaidya-de Sitter metric takes precisely the form \eqref{kot2} with the mass parameter replaced by an arbitrary function of $v$:
\be\label{VdS}
ds^2 = -\left(1-\frac{2m(v)}{r}-\frac{\Lambda}{3} r^2\right) dv^2 +2dv dr +r^2 d\Omega^2 
\ee
where $d\Omega^2$ is the standard metric of the unit round sphere, $r$ is the areal coordinate (so that round spheres with $v$ and $r$ constant have area $4\pi r^2$) and the range of coordinates is $v\in(-\infty,\infty)$ and $r\in (0,\infty)$ (or $r\in (-\infty,0)$). 
The metric \eqref{VdS} is a solution of the Einstein field equations with cosmological constant for an energy-momentum tensor of null radiation
$$
T_{\mu\nu}= \frac{2}{r^2}\frac{dm}{dv}k_\mu k_\nu
$$
where the future pointing null one-form $k_\mu$ and vector field $k^\mu$ are given respectively by
$$
\mathbf{k}=-d v, \hspace{1cm} \vec k =-\frac{\partial}{\partial r}.
$$
Thus, the massless particles of the `null dust' propagate along the null hypersurfaces $v=$ const.\ towards decreasing values of $r$, that is to say, towards round spheres of smaller areas. The dominant energy condition is satisfied whenever $m(v)$ is a non-decreasing function everywhere
\be\label{mdot}
\frac{dm}{dv}\geq 0
\ee
which I assume from now on. I will also assume $m\geq 0$ everywhere.

Kuroda \cite{K} proved that, under the above assumptions, a naked singularity would form in the Vaidya spacetime (with $\Lambda =0$)  if the mass function initially increases slowly, that is if  $m(v\rightarrow 0^+)/v\leq 1/16$ --see also \cite{FST,BeS}. This limit was confirmed for the Vaidya-de Sitter case in \cite{WM}, and thus for simplicity I am going to assume\footnote{If this condition does not hold, the main conclusions do not change: absence of $\scri^+$ and event horizon. The only difference will be the existence of another, null and locally naked, singularity in addition to the universal one in the future.}
\be\label{precond}
\lim_{v\rightarrow 0^+} \frac{m(v)}{v} > \frac{1}{16} .
\ee

The hypersurfaces $r=$const.\ have a normal one-form that satisfies
$$g^{\mu\nu}\partial_\mu r \partial_\nu r = 1-\frac{2m(v)}{r}-\frac{\Lambda}{3} r^2$$
so that they are always spacelike for large values of $r$ and for $r\longrightarrow 0$ if $m>0$. In those regions $r$ is a time coordinate. Those hypersurfaces can also be timelike if there are regions where the function above is positive. Fixing $v$, this can happen only if
\be\label{cond}
\Lambda < \frac{1}{9m^2(v)}
\ee
at that $v$. In particular this is always the case for $v$ such that $m(v)=0$. If condition \eqref{cond} holds at a given $v$, then there are two values of $r$, that I denote by $r_-(v)$ and $r_+(v)$, such that the round spheres with those values of $r$ at that value of $v$ are marginally trapped. It is easily seen that they satisfy
$$0<r_-(v)\leq \frac{1}{\sqrt{\Lambda}}\leq r_+(v)<\frac{3}{\sqrt{\Lambda}}$$
and that $r_-(v)$ increases, while $r_+(v)$ decreases, with $v$, that is, as $m(v)$ increases. Equality here is only possible if there exists a value $\bar v$ of $v$ such that
\be \label{vbar}
m (\bar v) = \frac{1}{3\sqrt{\Lambda}} .
\ee
The two hypersurfaces defined by $r=r_\pm(v)$ have a normal one-form given by
\be
\left( 1-\Lambda r_\pm^2(v) \right) dr -2 \frac{dm}{dv} dv
\ee
whose norm is \cite{BeS}
$$-4\frac{dm}{dv}(1-\Lambda r_\pm^2(v) )$$
so that these marginally trapped spheres pile up to form a {\em spacelike} marginally trapped tube (MTT) and a {\em timelike} MTT given by $r=r_-(v)$ and $r=r_+(v)$, respectively. Hence, the former is a dynamical horizon and the latter a timelike membrane, see \cite{AG,AK,BeS,Booth,S} for definitions.
I will denote by DH the former and by TM the latter. (Of course, in open regions where $m(v)$ is a --non-zero-- constant, they become null Killing horizons as the metric is Kottler there). Observe that, in case the value $\bar v$ in \eqref{vbar} exists, then 
$$
r_- (\bar v) =r_+ (\bar v) = \frac{1}{\sqrt{\Lambda}}
$$
so that DH and TM merge at the special round sphere defined by $v=\bar v$ and $r=1/\sqrt{\Lambda}$, and they both become null and tangent to the $v=\bar v$ null hypersurface there. This special round sphere has precisely the maximum area $4\pi/\Lambda$.

\section{First type of models}\label{sec:first}
\begin{figure}[h!]
\includegraphics[width=14cm]{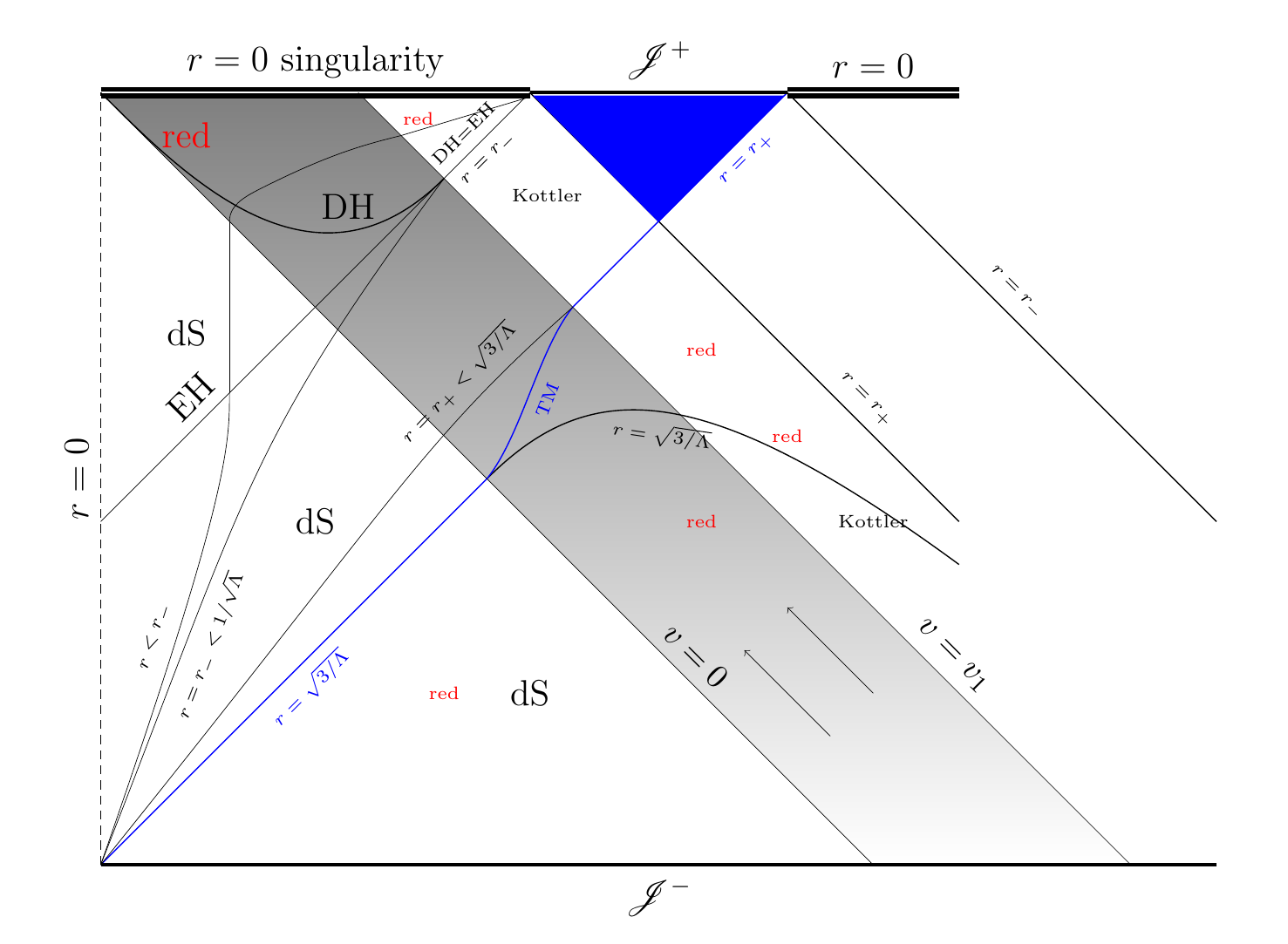}
\caption{{\footnotesize Conformal diagram of the Vaidya-de Sitter metric with $9 \mu^2 \Lambda <1$. Radial null geodesics are at 45$^o$ and future is upwards. Each point in the diagram represents a round sphere of area $4\pi r^2$ except the origin of coordinates on the left with $r=0$. Around this centre the metric is originally a portion of de Sitter, but at $v=0$ null matter collapses spherically towards that centre until the advanced time $v=v_1$. This is shown by the shadowed zone in the diagram, with the arrows pointing in the direction of propagation of the matter. The collapse produces a singularity and the appearance of future-trapped surfaces enclosed by an apparent horizon DH, which is a dynamical horizon foliated by marginally future-trapped round spheres with $r=r_-(v)$ (see main text). At $v=v_1$ one has $m(v_1)=\mu$, and the resulting spacetime to the right of $v=v_1$ is a BH of Kottler type and mass parameter $\mu$ , as indicated. Therefore, there is also a cosmological horizon at $r=r_+(v_1)$ and a blue region with past-trapped spheres to the past of future infinite $\scri^+$. The DH merges with the event horizon EH at the sphere $v=v_1$ with $r=r_-(v_1)$. In the figure $r_\pm (v_1)$ are simply represented by $r_\pm$, and they correspond to the two values of $r$ at the Killing horizons of the Kottler part. The original Killing horizon with $r=\sqrt{3/\Lambda}$ in dS becomes a timelike membrane TM with $r=r_+(v)$ (see main text) when it is crossed by the in-falling matter until the inflow terminates, where it becomes the Killing horizon $r=r_+$ of the Kottler metric corresponding to the mass parameter $\mu$. The corresponding hypersurfaces with $r=r_\pm (v_1)$ in the dS and shadowed regions are also drawn. The hypersurface $r=\sqrt{3/\Lambda}$ is spacelike for $v>0$ as indicated. The hypersurfaces with constant $r$ that cross the DH become null there and then spacelike, as the one shown with $r<r_-$. The round spheres are future trapped in the zone above DH, and in the zone to the right of the blue line --which is an MTT with two Killing horizon portions-- and below the $r=r_+$ on the Kottler part. Not to overwhelm the picture, this is simply indicated with several `red' words. 
The metric is extendible towards the right, and the analytical extension is that shown in the Appendix for the Kottler metric.}}
\label{fig:VdS1}
\end{figure}

The first type of models I am going to consider are defined by imploding null dust into an empty de Sitter universe. Thus, the mass function $m(v)$ is assumed to vanish at initial values of $v$. At a given advanced time (say $v=0$) $m(v)$ starts to increase until eventually reaches its maximum value $\mu$, say at $v=v_1$:
\be
m(v\leq 0)=0, \hspace{1cm} m(v\geq v_1)=\mu
\ee
where the condition \eqref{mdot} holds for $v\in (0,v_1)$, and \eqref{precond} is enforced too.

There are two possibilities to be considered, depending on whether $9\mu^2 \Lambda <1$ or not. The case usually analyzed in the literature has $9\mu^2 \Lambda <1$, see e.g. \cite{ABK}, especially concerning BH evaporation \cite{M,Mallet1,H,ZYR} because it leads to a standard Kottler (or Schwarzschild-de Sitter) black hole. The conformal diagram is presented in Figure \ref{fig:VdS1}.

\begin{figure}[h!]
\includegraphics[width=16cm]{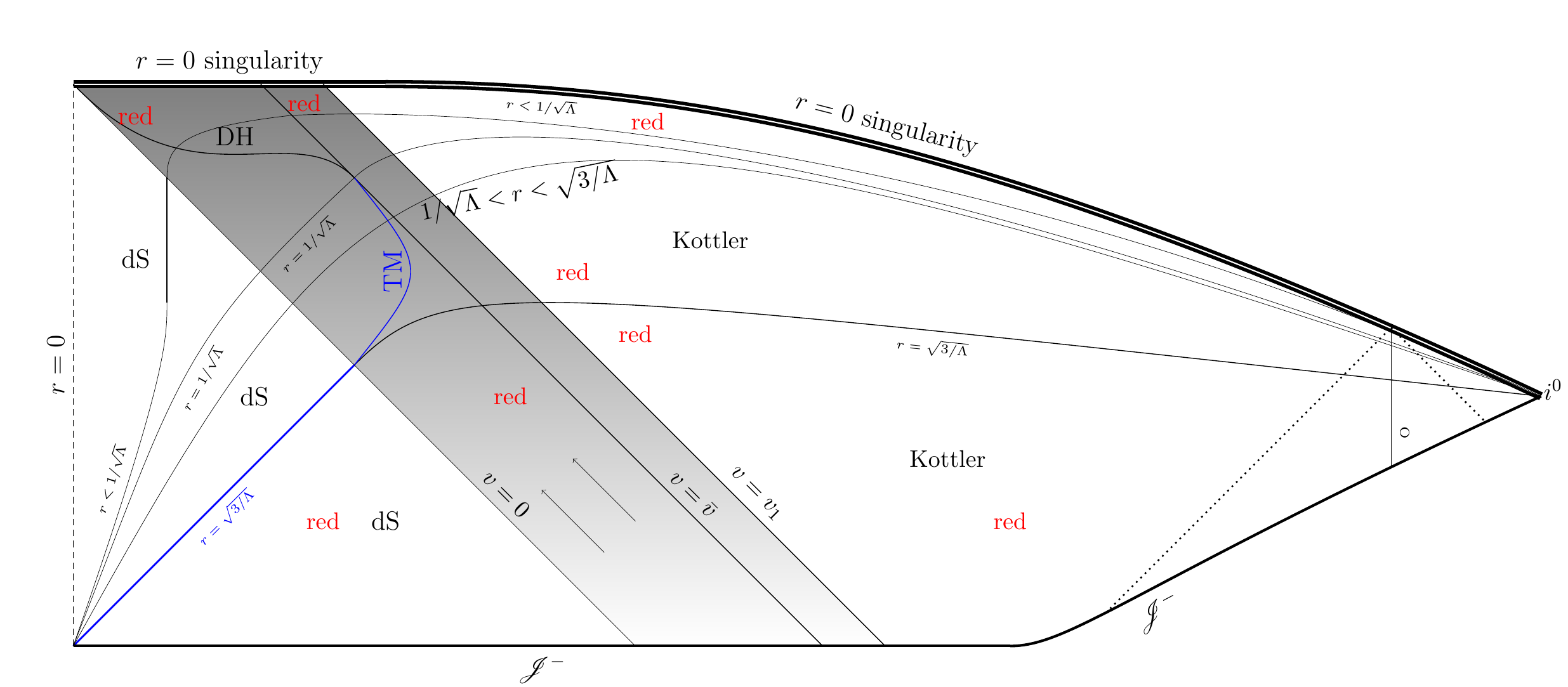}
\caption{{\footnotesize Conformal diagram of the Vaidya-de Sitter metric with $9 \mu^2 \Lambda >1$, same conventions as in the previous figure. The metric is originally a portion of de Sitter and at $v=0$ null matter coming from past null infinity collapses spherically towards the regular centre at $r=0$ until the advanced time $v=v_1$, where $m(v_1)=\mu$. This is shown by the shadowed zone in the diagram, with the arrows pointing in the direction of propagation of the matter. To the right of $v=v_1$ the spacetime is of Kottler type with mass parameter $\mu$, that is, of the type shown in Figure \ref{fig:kot2}. Therefore, there is no cosmological horizon (no blue region with past trapped spheres) nor future infinity $\scri^+$ in this situation. As in the previous case, the collapse produces a singularity and the appearance of future trapped surfaces enclosed by an apparent horizon AH, which is a dynamical horizon foliated by marginally future-trapped round spheres with $r=r_-(v)$. However, at $v=\bar v$ one reaches the extreme value $m(\bar v) =1/(3\sqrt{\Lambda})$, hence the AH must end up there somehow. The original Killing horizon with $r=\sqrt{3/\Lambda}$ in dS becomes a timelike MTT with $r=r_+(v)$ (see main text) when it encounters the in-falling matter and then the area of their marginally trapped spheres decreases until asymptotically tends to $4\pi/\Lambda$ ($r=1/\sqrt{\Lambda}$) and to a merging with AH. Both AH and MTT become null at the crucial, marginally trapped, round sphere defined by $v=\bar v$ and $r=r_\pm(\bar v) =1/\sqrt{\Lambda}$. Some hypersurfaces with constant $r$  are also shown, they are timelike to the left of MTT (and below AH), become null at MTT and AH, and are spacelike to the right of the MTT (and above AH). All of them reach spacelike infinity $i^0$ asymptotically. Future trapped round spheres are abundant as they are un-trapped only in the zone to the left of MTT and below AH. This is a small portion of the spacetime, because a spacelike disk centered at $r=0$ and reaching up to the MTT has a finite volume, while spacelike cylinders starting anywhere in the diagram with $r>0$ and reaching $i^0$ have an infinite volume. Many observers, such as the $O$ showed, will never be able to see (or be influenced by) the matter creating the strong gravitational field, but they feel the latter. The $O$-particle horizon enclosing its complete past is shown by the dotted lines. Every possible observer in this spacetime ends up  at the future singularity. The metric is globally hyperbolic and inextendible.}}
\label{fig:VdS2}
\end{figure}

However, in this paper I want to consider the other possibility, that is, when the final mass parameter satisfies $9\mu^2 \Lambda >1$. Therefore, now the value $\bar v$ in \eqref{vbar} does exist. In this situation a spherically symmetric dynamical horizon DH emerges and the area of the marginally trapped surfaces foliating DH increases with $m(v)$ until it reaches its maximum possible value \eqref{lim} at $v=\bar v$, but $m(v)$ keeps growing beyond that value until it reaches $m(v_1)=\mu > m(\bar v)$. This leads to the absence of future null infinity $\scri^+$ and thereby to the nonexistence of any event horizon EH. What was going to become a BH actually grows up `too much'  and ends up swallowing the entire spacetime that becomes a contracting universe of type \eqref{kot3} outside the matter. Now the singularity is universal and every single possible observer or photon will inevitably end up there in finite time (or affine parameter). The whole thing is explained, and perhaps better understood, in the corresponding conformal diagram of Figure \ref{fig:VdS2}.

The spacelike dynamical horizon DH and the timelike membrane TM merge and become null at the crucial, marginally trapped, round sphere defined by $v=\bar v$ and $r=r_\pm(\bar v) =1/\sqrt{\Lambda}$. This produces a single hypersurface with signature-changing character that has been called a ``future holographic screen'' in \cite{BE,BE1} because it satisfies an area law: starting from the upper left corner in the diagram of Figure \ref{fig:VdS2}, and following this hypersurface all the way until it reaches the dS part of the spacetime, the area of the foliating marginally trapped spheres is monotonically increasing.

The null hypersurface $v=\bar v$ is a past horizon for the region with marginally trapped round spheres, and any event with $v>\bar v$ is unable to influence any such MTS. This region containing marginally trapped round spheres is actually a small (finite) portion of the entire spacetime, as can be proven by computing the volume of spacelike slices contained in the appropriate regions. On the one hand, the volume of spacelike spherically symmetric hypersurfaces orthogonal to the $r=$const.\ hypersurfaces contained in the region with $1-\frac{2m(v)}{r}-\frac{\Lambda}{3} r^2 >0$ have a volume
$$
4\pi \int_0^{r_\pm(v)} \frac{r^2}{\sqrt{1-\frac{2m(V(r))}{r} -\frac{\Lambda}{3} r^2}} dr =\mbox{finite}
$$
where $v=V(r)$ is the function defining these hypersurfaces. On the other hand, the spacelike spherically symmetric hypersurfaces $r=$const.\ in the Kottler region $1-\frac{2\mu}{r}-\frac{\Lambda}{3} r^2 <0$ have an infinite volume
$$
4\pi r^2  \sqrt{\frac{\Lambda}{3} r^2+\frac{2\mu}{r} -1} \int_{v_1}^\infty dv = \infty . 
$$
\begin{figure}[h!]
\includegraphics[width=16cm]{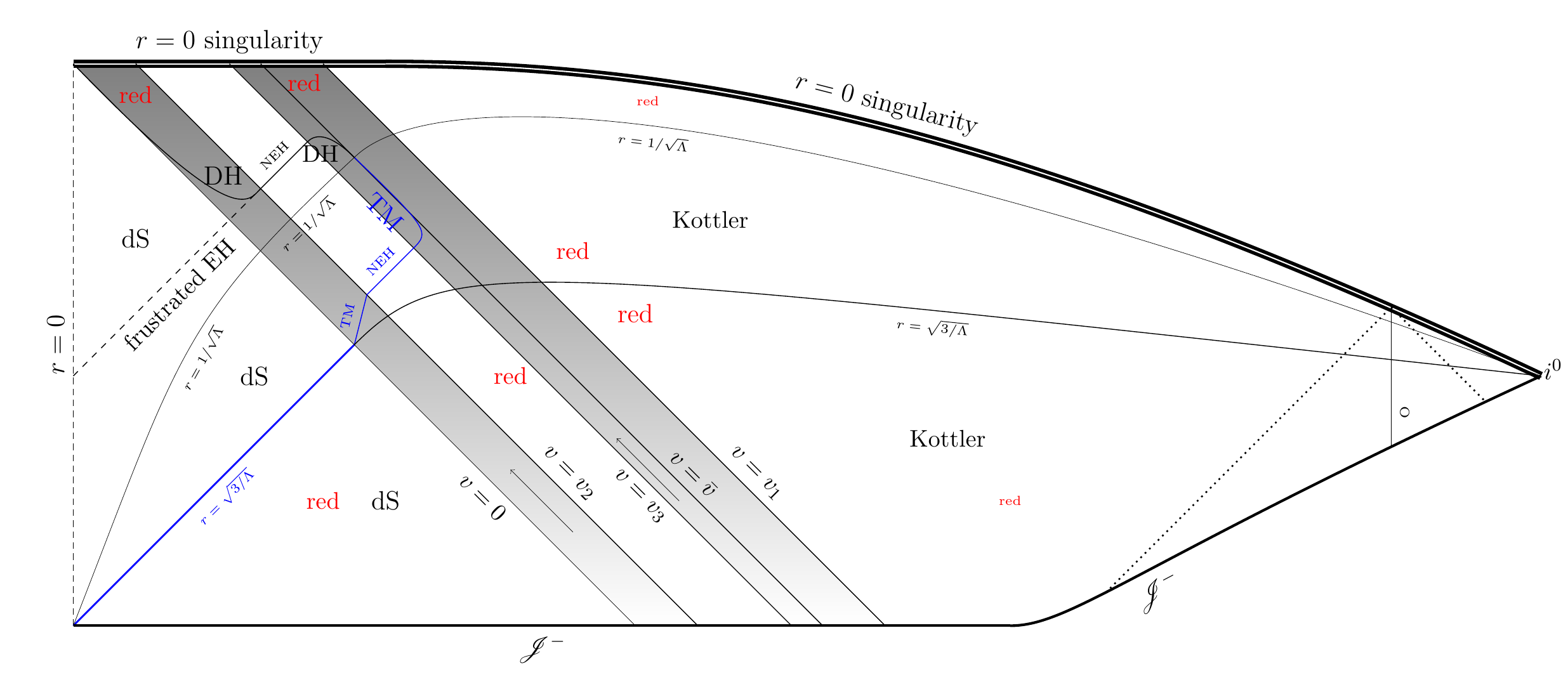}
\caption{{\footnotesize Another conformal diagram of the Vaidya-de Sitter metric with $9 \mu^2 \Lambda >1$, same conventions as in the previous figure. Now the flow of incoming radiation stops at $v=v_2$ until $v=v_3$, where more matter comes in until the advanced time $v=v_1$, where $m(v_1)=\mu$. The two zones with matter are shown by the shadowed strips in the diagram, with the arrows pointing in the direction of propagation of the matter. To the right of $v=v_1$ the spacetime is of Kottler type with mass parameter $\mu$, that is, the same portion as in Figure \ref{fig:VdS2}. Therefore, the main interesting features as in the previous figure remain: there is no cosmological horizon, no future infinity $\scri^+$, and no EH. All features for $v<v_2$ and for $v\geq v_3$ are the same as in the spacetime of Figure \ref{fig:VdS2}. However, the key difference here is the existence of the region $v_2 < v < v_3$ where there is no matter and the metric is Kottler with two non-expanding (actually Killing) horizons denoted by NEH. One should observe that the region with $v< v_3$ is identical with a portion of the spacetime in Figure \ref{fig:VdS1}, where an EH had formed and a BH is settled down. Hence, one can argue that the spacetime with $v<v_3$ is, in that region, a temporary BH in equilibrium with mass parameter $m(v_2)$. However, the null hypersurface destined to be the EH of such a BH in equilibrium never becomes an actual event horizon, here marked by `frustrated EH', as a second flow of matter turns the spacetime into another ultra-massive universe.}}
\label{fig:VdS3}
\end{figure}

It is remarkable that the absence of future null infinity arises precisely because there is a positive cosmological constant. As proven in \cite{K}, the non-existence of $\scri^+$ leading to the absence of the EH requires, for the Vaidya metric (that is, with $\Lambda =0$), that 
$$
\lim_{v\rightarrow \infty} \frac{m(v)}{v} > \frac{1}{16}.
$$
In plain words, when $\Lambda =0$ one needs a very large {\em infinite} total mass. However, the existence of $\Lambda > 0$ changes this drastically and any finite $m(v)$ larger than $3/\sqrt{\Lambda}$ eliminates $\scri^+$ and the EH. As the spacetime was locally creating a BH for a period of advanced time, these are somehow {\em frustrated black holes}, but the frustration arises simply because there are no observers reaching infinity, so the would-be BH ends up being a victim of its own success as its mass increases beyond the acceptable limit for the area of MTS. Therefore, recalling that the word `ultra' comes from the Greek `beyond', I think an acceptable name for these type of models is {\em ultra-massive spacetimes}. 

To make the features of these models more explicit and, perhaps, surprising, let us consider a mass function with the following properties
$$
m(v\leq 0)=0, \hspace{1cm} m(v_2<v < v_3) =\mu_0 < \frac{3}{\sqrt{\Lambda}}, \hspace{1cm} m(v\geq v_1)=\mu >\frac{3}{\sqrt{\Lambda}}
$$
with  $0<v_2 <v_3<\bar v <v_1$ and $\mu_0$ a constant less than the critical value, while keeping \eqref{mdot} everywhere. This describes a situation identical to the first model, i.e. the creation of a BH of mass $\mu_0$, for all $v< v_3$, as the model settles down to a would-be BH with potential EH corresponding to the Killing horizon with $r=r_-(\mu_0)$ of the Kottler metric with mass $\mu_0$. Nevertheless, this will eventually become a frustrated EH due to the extra matter that falls into the would-be BH after $v=v_3$. Eventually, the BH never forms, again victim of its own success in accumulating matter. The corresponding conformal diagram is presented in Figure \ref{fig:VdS3}. 
I would like to remark that the zone $v_2<v<v_3$ can be made extremely large, so that such a `frustrated BH' can look like a real BH in equilibrium for a period of time that can be taken as large as desired.

\begin{figure}[h!]
\includegraphics[width=13cm]{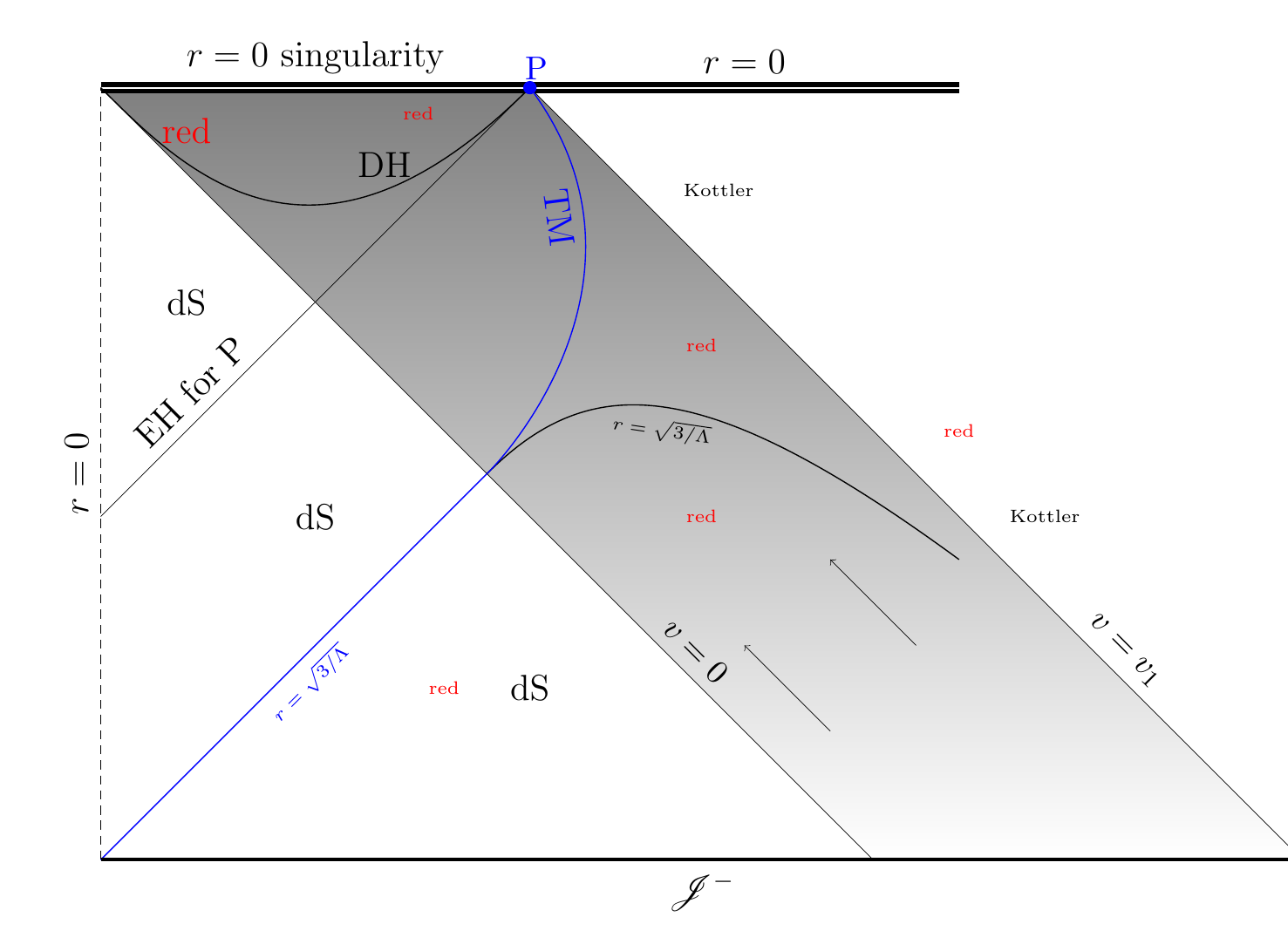}
\caption{{\footnotesize This is the scematic diagram for the extreme case $9 \mu^2 \Lambda =1$. Now the flow of incoming radiation stops at $v=v_1$ with $m(v_1)=\mu =1/(3\sqrt{\Lambda})$. To the right of $v=v_1$ the spacetime is of extreme Kottler type, as in Figure \ref{fig:kotext} in the Appendix. The TM and DH now tend to merge at infinity, approaching $r=1/\sqrt{\Lambda}$ asymptotically. The `point' P is at infinity, and some (but not all) observers may reach there. For this type of observers there remains a caricature of the EH.}}
\label{fig:VdSext}
\end{figure}

There is a limiting case when $v_1=\bar v$, that is to say, $m(\bar v) =m(v_1)=\mu =3/\sqrt{\Lambda}$. The corresponding conformal diagram can be easily drawn, by taking the required part of the extreme Kottler metric joined to a version of Figure \ref{fig:VdS2} with $v_1 =\bar v$, to the left of $\bar v$. In this case the DH and TM both tend to merge at future infinity, arriving at the infinity `point P' of the type shown in Figure \ref{fig:kotext} of the Appendix. In this case, P is the only remaining vestige of the existence of future infinity, and there are some very special observers that can actually reach there. Therefore, for this special class of observers one can still define an EH. I have included the conformal diagram for compelteness in Figure \ref{fig:VdSext}.

\section{Second type of models}\label{sec:second}
The second type of models I am going to consider consist of BHs already formed (or better said in formation) by stellar collapse that, after having settled down to equilibrium, receive further matter that makes them grow beyond the limit \eqref{lim}. For illustration purposes I am going to use the generalization of the Oppenheimer-Snyder collapse \cite{OS,BJS} to the case with $\Lambda >0$ analyzed in \cite{GV,N,L,MS} years ago. Nevertheless, the construction of the models work for any other collapse that produces a Kottler (Schwarzschild-de Sitter) BH, such as those studied in \cite{DJCJ}.

\begin{figure}[h!]
\includegraphics[width=14cm]{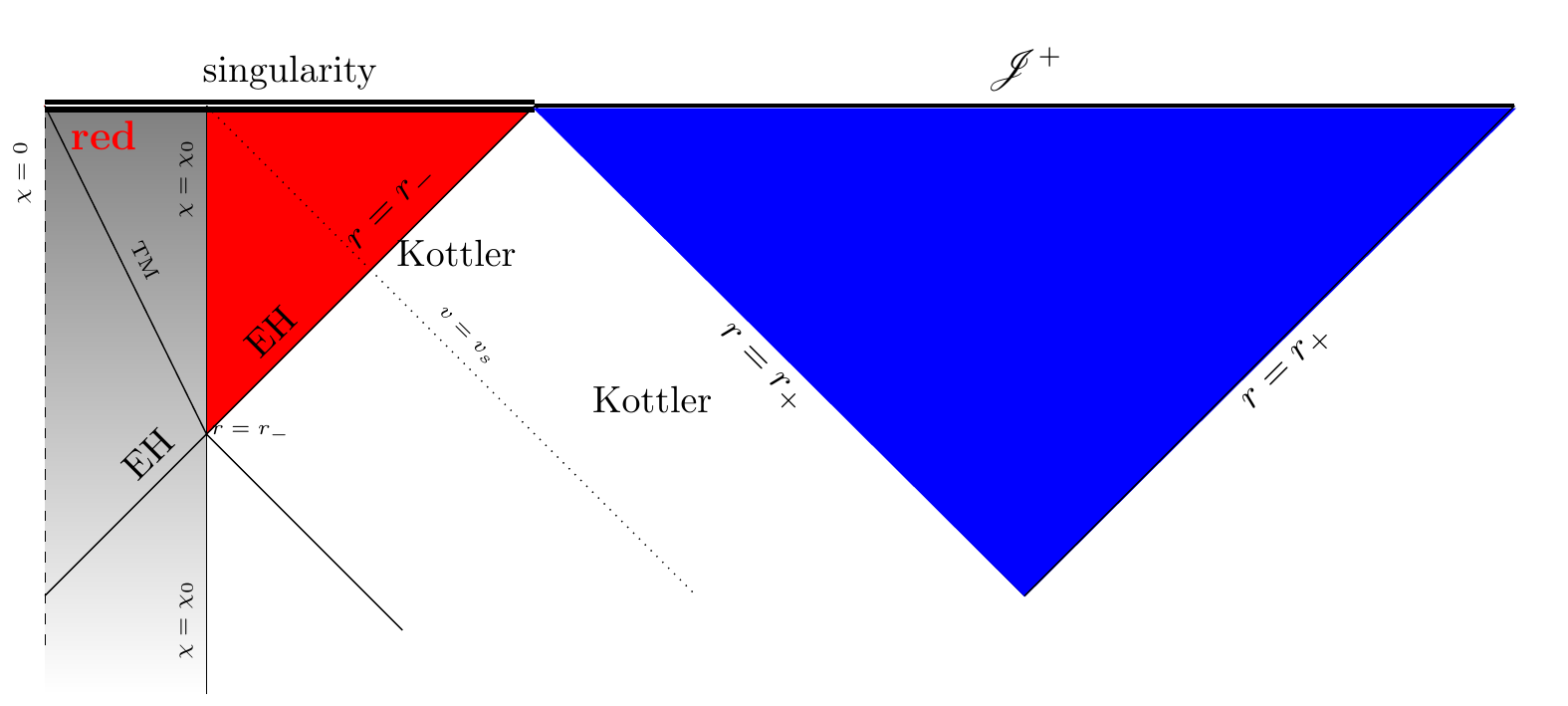}
\caption{\footnotesize{Conformal diagram of the formation of a BH by collapsing homogeneous dust in the presence of $\Lambda >0$, same conventions as before (now the centre of the dust is the line $\chi =0$). One starts with initial conditions at an instant of time symmetry, a spacelike hypersurface with initial zero velocity $da/d\tau =0$ in the dust region. This dust portion is represented here by the shadowed zone. The timelike hypersurface $\chi=\chi_0$ is the boundary of the collapsing dust and the spacetime is matched at this hypersurface to the unique spherically symmetric vacuum exterior, which is Kottler metric with mass parameter $m$ such that \eqref{am}, or equivalently \eqref{am1}, holds. The collapse leads to the existence of a timelike TM foliated by marginally trapped round spheres, which the dust surface intersects at the time that the dust cloud has an area $4\pi r_-^2$. As usual, assuming that $9m^2 \Lambda <1$ there is also a cosmological horizon with $r=r_+$ and corresponding future null infinity $\scri^+$. Hence there is a BH with EH at $r=r_-$ in the Kottler part, as shown. The marginally trapped round spheres foliating the TM are not spatially stable in the sense of \cite{AMS}, the stable ones --that necessarily exist \cite{E,AM}-- are those foliating the EH in the Kottler region with area $4\pi r_-^2$. All round spheres in the red zones are future trapped, including the part of the dust cloud above TM, indicated here with the word `red' not to overlap with the shadow of the dust. The blue region contains past-trapped round spheres. The metric is extendible towards the right. I have represented the null hypersurface $v=v_s$, defined as the limit of advanced times $v$ that reach the dust cloud. Therefore, at any $v>v_s$ one can match this spacetime to a Vaidya-dS metric in the same fashion as in Figure \ref{fig:VdS3} to produce another ultra-massive spacetime where the EH and the BH disappear. This is shown in the next diagram, Figure \ref{fig:OSVdS}.}}
\label{fig:OSdS}
\end{figure}

The metric is described by the matching of the Kottler metric \eqref{kot} to a closed Robertson-Walker metric ($0<\chi<\pi$)
\be
ds^2 =-d\tau^2 +a^2(\tau) \left(d\chi ^2 +\sin^2\chi d\Omega^2 \right)
\ee
where the scale factor solves the Fridman-Lema\^\i tre equation for dust (i.e., pressure $p=0$)
\be\label{FLeq}
\left(\frac{da}{d\tau}\right)^2 =\frac{a_m}{a} +\frac{\Lambda}{3} a^2 -1 .
\ee
Here $a_m$ is a constant that represents the minimum value of the dust mass density and which,  via the matching, can be related to the exterior (Kottler) constant mass parameter $m$ by
\be\label{am}
a_m \sin^3\chi_0 =2m
\ee
and the matching hypersurface is defined by
$$
\chi =\chi_0 , \hspace{1cm} r=r_\Sigma = a(\tau) \sin\chi_0
$$
in the interior and exterior parts, respectively. The constant $\chi_0$ is choosable in principle, and one can easily check on using \eqref{am} the relation
\be\label{am1}
\frac{4\pi}{3} \rho r_\Sigma^3 = \frac{c^2}{G} m (=M)
\ee
where $\rho$ is the mass density of the dust cloud. The righthand side of equation \eqref{FLeq} never vanishes if $9m^2 \Lambda >1$, connecting with the conditions on the exterior for the absence of Killing horizons, see \cite{N} for details.\footnote{This situation was also considered in \cite{MS} in their section V, and the exterior is of type \eqref{kot3} leading to a conformal diagram which has a portion of the Figure \ref{fig:kot2} as the exterior part. However, this was not presented in \cite{MS}, as the authors chose to place a second collapsing dust to the right of the diagram---see their figure 10. Nevertheless, these models do not have any MTS anywhere, and therefore they are not of interest in the present discussion.} However, I am going to consider the other possibility in which an MTT (of TM type) arises and a BH is formed. Thus choosing $9m^2 \Lambda <1$ there always exist values of $\chi_0 <\pi/2$ such that there is a bouncing time where $d a/d\tau=0$. This time is usually taken as the initial time of the dust collapse \cite{OS,MS}. The collapse leads to a BH of Kottler type and total mass $M$. The conformal diagram is presented in Figure \ref{fig:OSdS}, see also \cite{N,MS}.

\begin{figure}[h!]
\includegraphics[width=16cm]{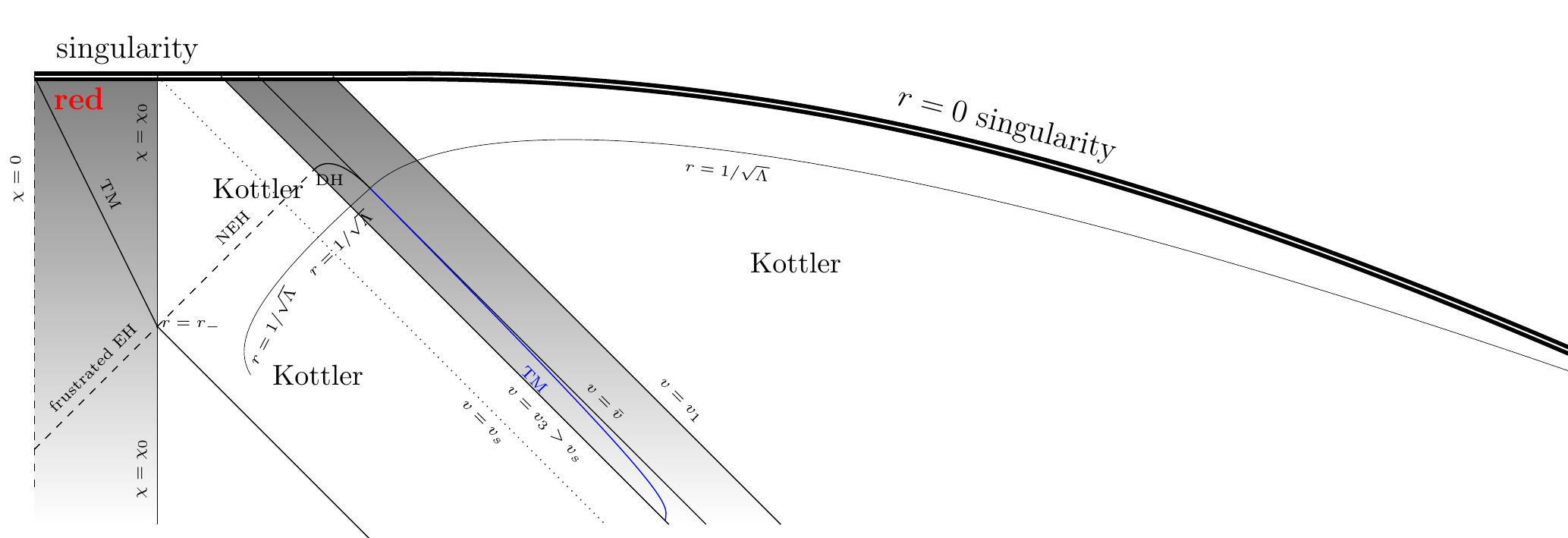}
\caption{\footnotesize{Conformal diagram of a spacetime where collapsing homogeneous dust creates a temporary would-be BH that eventually is frustrated by the reception of extra matter in the form of null dust, such that $9\mu^2 \Lambda >1$. The dust cloud is represented by the shadowed zone to the left of the timelike hypersurface $\chi =\chi_0$ that represents the surface of the collapsing star.  All conventions are as before. The part of this diagram to the left of $v=v_3> v_s$ is identical to that part of the diagram in Figure \ref{fig:OSdS}, while the portion to the right of $v=v_3 >v_s$ is the same as the corresponding portion in Figure \ref{fig:VdS3}. Hence, this is just another example of a frustrated BH, that might have been settled down and in equilibrium for very long ages, but eventually transforms into an ultra-massive spacetime with no $\scri^+$ and no EH, producing a universal future singularity.}}
\label{fig:OSVdS}
\end{figure}

Now, the idea is to throw matter into this `already formed' black hole in order to create an ultra-massive spacetime, so that no BH remains. To do that, one simply has to throw enough matter into the BH. This can be easily accomplished by using, as in previous cases, the Vaidya-dS metric starting at any $v> v_s$, where $v_s$ is defined as the limit of the advance times $v$ that reach the dust cloud (so that the entire matching is performed in the Kottler part, for simplicity). If the final total mass parameter $\mu$ is larger than $1/(3\sqrt{\Lambda})$ we again encounter a situation where something that looks like a BH in equilibrium for a long time, and was {\em formed by  stellar collapse}, eventually becomes an ultra-massive spacetime with no $\scri^+$ and no event horizon. This is represented in Figure \ref{fig:OSVdS}.

\section{Discussion}\label{sec:discussion}
First of all, I would like to remark that, despite the fact that these are idealized simple models, the conclusion is robust in spherical symmetry\footnote{This also follows from the uniqueness results of the Oppenheimer-Snyder-like models in spherical symmetry, which themselves follow, via the idea of complementary matchings \cite{FST}, from the uniqueness of the Einstein-Straus vacuoles, see \cite{MMV} for details.}: if one tries to increase the area of a MTS beyond the limit \eqref{lim} by throwing matter into its interior the outcome will be the end of the stable MTSs as there will be one with the maximum area \eqref{lim}. This entails the dematerialization of the EH and of $\scri^+$, implying a general collapse into a future universal curvature singularity. The best way to understand that this is a general result is to recall the results by Friedrich, stating that solutions of the Einstein field equations are uniquely determined by initial data on $\scri^-$ \cite{Frie0}.  Therefore, if the initial data contains a portion of $\scri^-$ with vacuum initial data that entail a total mass larger than $1/(3\sqrt{\Lambda})$, then we know for sure that a portion of the conformal diagram is provided by a portion of the Kottler metric given in Figure \ref{fig:kot2} in the Appendix, thus leading to the future singularity.  I conjecture that the conclusion still holds without spherical symmetry, for instance, using the results in \cite{BP,PS} where the Robinson-Trautman metrics with $\Lambda$ are seen to possess properties similar to Vaidya-dS, and actually they all approach Vaidya-dS asymptotically.

As we have seen, the limit \eqref{lim} is not violated in any of the models, even when increasing the total mass of the spacetime. Somehow, General Relativity is prepared to accept as much mass as one can imagine, nevertheless {\em spatially} stable marginally trapped surfaces cannot increase its area indefinitely if there is a positive $\Lambda$. They simply approach an MTS with the maximum area that ceases to be stable in any possible spacelike direction. This can be better understood by noticing that the dynamical horizon DH, which is foliated by marginally trapped round spheres that are always stable in some spacelike outward directions\footnote{The outward direction here is the null future direction with vanishing expansion, or equivalently the direction into which the (null) mean curvature vector of the MTS points \cite{S,BeS}}, eventually merges with a TM where the marginally trapped spheres are not stable in any spacelike direction; and they merge becoming tangential to a null hypersurface. Therefore, the special round sphere where they merge cannot be deformed outwardly in any non-timelike direction without becoming a (weakly) trapped surface.

It must be observed that the results of persistence of stable MTS \cite{AMMS} are not in conflict with the models we have presented because they require the existence of an exterior untrapped barrier \cite{AM,E}. This is a completely external surface `enclosing' the MTS and joinable to it by a spacelike hypersurface, thus leading to stability within this hypersurface, and this is precisely what is missing at the special MTS where DH and TM merge ---this is also true for all MTS to the past of the distinguished one--, as there are no untrapped external spheres whatsoever.

An important puzzling question that arises is that of BH evaporation via Hawking radiation. The usual picture cannot be applied here as there is no EH defining the black hole. Of course, it has long been argued that Hawking radiation may have a different origin \cite{Kod} such as dynamical horizons or marginally trapped tubes of the type DH and TM \cite{Haj}. One can even argue that some kind of radiation can be associated to {\it any} MTS \cite{ST}. However, in the ultra-massive models herein presented, the question is where does any such radiation go. There is no infinity that allows the system to radiate (lose) energy away, and thus the already infinite curvatures at the singularity will become even larger if some energy arrives there from somewhere else. How quantum gravity might resolve this puzzle is uncertain. These results have also some implications on how to deal with BHs mergers and how to use numerical relativity to describe them. Because there seems to be a limit for the merger of apparent horizons and, if this limit is surpassed, outwardly stable MTSs simply fail to exist: no numerical code will ever find them.

Of course, there is also the query of how much mass is necessary to produce such ultra-massive spacetimes, and this depends on the value of the cosmological constant. If we accept the value that arises from the observed accelerated expansion of the visible Universe, which is about \cite{Planck}
$$
\Lambda \simeq 1.1 \times 10^{-52} \mbox{m}^{-2} ,
$$
the limit \eqref{lim} requires a gravitational radius $2m$ that should be greater than
$$
6.4 \times 10^{25} \, \mbox{m}
$$
and this translates into a total mass of about
$$
2.2\times 10^{22} M_\odot\sim 4.3 \times 10^{52} \mbox{Kg} .
$$
The estimated total mass of the observable universe now is about 
$$
8.8\times 10^{52} - 1.0 \times 10^{54} \mbox{Kg}
$$
so that one would need up to one-third of the total observed mass {\em now} to produce such ultra-massive objects. It does not seem they are going to be seen in the forseeable future! Still, the total mass of the entire Universe may well be much larger than that, hence  these objects might be real somewhere, some time. And, in any case, there is a question of principle: if they are permitted by the theory for any value of $\Lambda >0$, what is the relevant physics behind them and how to deal with the universal singularity?

Finally, I would like to add a remark. The time reversals of ultra-massive spacetimes are also worth considering. One just has to look at the diagrams upside down, so that the future direction is reversed. Then, for instance, the time reversal of the model represented in Figure \ref{fig:VdS2} will describe a universal big-bang singularity in the past and an expanding Universe of locally rotationally symmetric (Kantowski-Sachs) type \cite{Exact} in the Kottler region, but the mass-energy creating the gravitational field is radiated away towards $\scri^+$ entirely, leaving behind a portion of de Sitter vacuum spacetime. And the model of Figure \ref{fig:OSVdS} will have two expanding regions coming from the big-bang singularity, one of them of FLRW type. This may lead to several interesting speculations.

\section*{Acknowledgments}
Research supported by the Basque Government grant number IT1628-22, and by Grant PID2021-123226NB-I00 funded by the Spanish MCIN/AEI/10.13039/501100011033 together with ``ERDF A way of making Europe'' . This research was carried out during a visiting professorship at Yukawa Institute for Theoretical Physics.

\begin{figure}[h!]
\includegraphics[width=10cm]{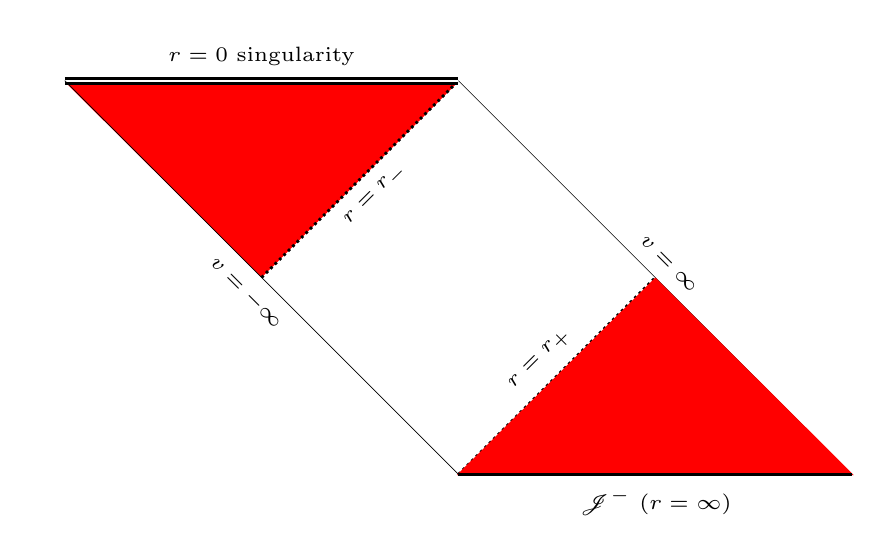}
\caption{{\footnotesize Conformal diagram of the Kottler metric covered by the coordinates in \eqref{kot2} when $9 m^2 \Lambda <1$. Radial null geodesics are at 45$^o$ and future is upwards. Each point in the diagram represents a round sphere of area $4\pi r^2$. The region near past infinity $\scri^-$ is (close to) de Sitter, while the region close to the future singularity is similar to that part of Schwarzschild. The spheres in the red regions are future-trapped, while the white zone between $r_+$ and $r_-$ is static --and thus free of compact trapped surfaces \cite{MaSe}. The dotted lines with $r=r_\pm$ are Killing horizons foliated by marginally trapped round spheres. The metric is extendible across $v=\pm \infty$, where the geodesics arrive with $r\longrightarrow r_\pm$ respectively. The analytical extension can be easily built by gluing copies of this patch and its time reversal in the appropriate way, as shown in Figure \ref{fig:kot1}.}}
\label{fig:kot}
\end{figure}

\section*{Appendix: The Kottler metrics}
The unique family of spherically symmetric vacuum solutions of the Einstein field equations --including a cosmological constant $\Lambda$-- is given by the Kottler (also known as Schwarzschild-de Sitter) metric \cite{Kot,Exact}
\be\label{kot}
ds^2 = -c^2 \left(1-\frac{2m}{r}-\frac{\Lambda}{3} r^2\right) dT^2 + \left(1-\frac{2m}{r}-\frac{\Lambda}{3} r^2\right)^{-1} dr^2 +r^2 d\Omega^2
\ee
where $d\Omega^2$ is the standard metric of the unit round sphere, $r$ is the areal coordinate and $T$ is a fourth coordinate with range in $(-\infty,\infty)$. The `mass parameter' is given by $m:=GM/c^2$ where $M$ is interpreted as the total mass generating the spacetime. When $m=0$ it reduces to a (static) portion of de Sitter (dS) spacetime. In all cases, when $r\rightarrow \infty$, the metric tends to dS.

\begin{figure}[h!]
\includegraphics[width=9cm]{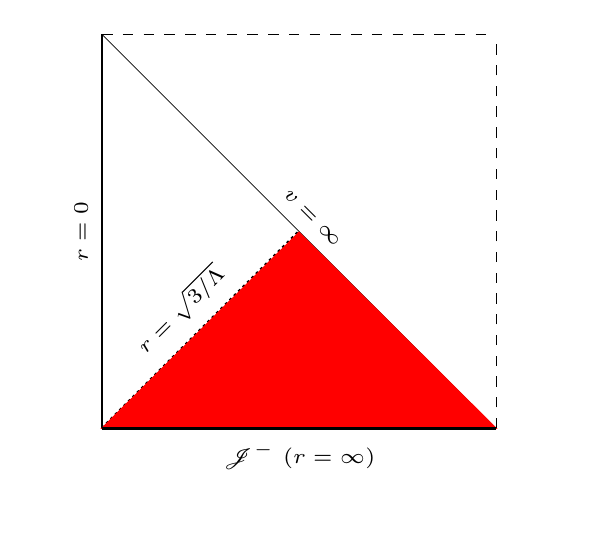}
\caption{{\footnotesize Conformal diagram of the portion of de Sitter spacetime covered by the coordinates in \eqref{kot2} with $m=0$, same conventions as in Fig.\ref{fig:kot} except that now $r=0$ is regular and can be seen as the origin of coordinates. The spheres in the red regions are future-trapped. The dotted line with $r=\sqrt{3/\Lambda}$ is a Killing horizon foliated by marginally trapped round spheres. The metric is extendible across $v= \infty$, where the geodesics arrive with $r\longrightarrow \sqrt{3/\Lambda}$. The analytical extension leads to the standard `square' diagram of de Sitter, shown here by the dashed lines. In the complete diagram the left ($r=0$) and right vertical lines represent the north and south pole of a 3-sphere. The topology of $\scri^-$ is, for the case of the portion covered by the coordinates in \eqref{kot2}, $\mathbb{R}\times \mathbb{S}^2$ --the two extremes of the segment representing $\scri^-$ are not part of the diagram. The slices of the spacetime have topology $\mathbb{R}^3$ if one includes the point $r=0$ at each instant of time.}}
\label{fig:dS}
\end{figure}

\begin{figure}[h!]
\includegraphics[width=13cm]{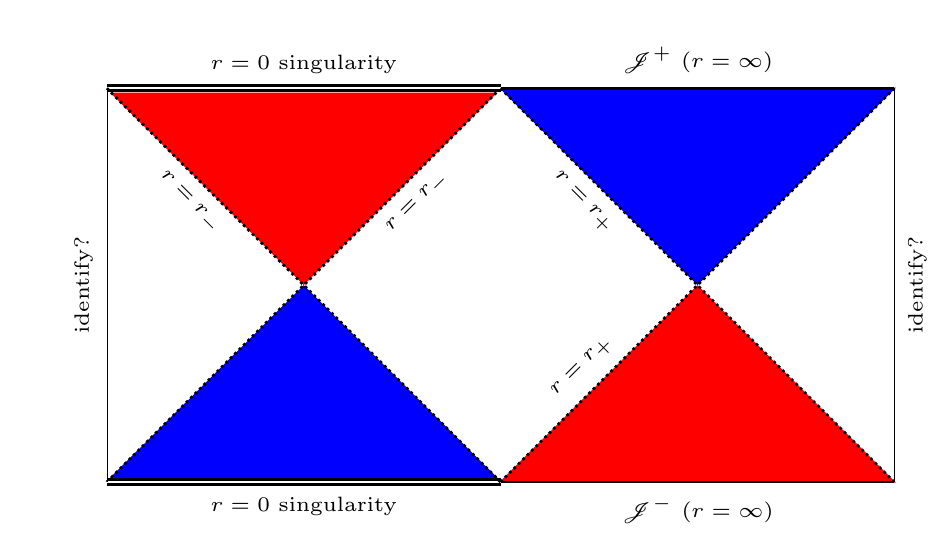}
\caption{{\footnotesize Conformal diagram of the extended Kottler metric with $9m^2 \Lambda <1$, same conventions as in Fig.\ref{fig:kot}. The spheres in the red regions are future-trapped, and those in the blue regions are past-trapped, while the white zones between $r_+$ and $r_-$ are static --and thus free of compact trapped surfaces. The dotted lines with $r=r_\pm$ are Killing horizons foliated by marginally trapped round spheres. The blue and red regions on the left thus represent white and black hole regions, respectively, with $r=r_-$ as the hole bifurcate horizon \cite{Wald}. The red and blue regions on the right represent past and future cosmological zones approaching, at past and future infinity respectively, de Sitter spacetime. The metric is analytically extendible towards the left and the right by just adding copies of the same diagram, leading to many, or infinite, BH regions.  The ``points'' where the singularities and infinities ``touch'' are not part of the diagram, and the topology of $\scri$ is $\mathbb{R}\times \mathbb{S}^2$. This is also the topology of space sections of the spacetime (say horizontal lines in the diagram). An alternative is to identify the left and right vertical lines, which will produce just one BH and one asymptotic region, then changing the topology of the space sections to $\mathbb{S}\times \mathbb{S}^2$.}}
\label{fig:kot1}
\end{figure}

 As is well known, when $\Lambda >0$ there are three different possibilities for the metric \eqref{kot} depending on whether $9m^2$ is greater, equal, or smaller than $1/\Lambda$. The standard case, which includes a static region similar to that of the Schwarzschild metric, requires 
$$
\Lambda < \frac{1}{9m^2}
$$
in which case the function $1-\frac{2m}{r}-\frac{\Lambda}{3} r^2$ has two positive zeros, $r_+$ and $r_-$ say. The particular values of $r_\pm$ can be found in \cite{LR,MS}. The two hypersurfaces defined by $r=r_\pm$ can be easily proven to define Killing horizons \cite{Wald} of the Killing vector $\partial_T$ through which the metric \eqref{kot} can be extended via the usual techniques. For instance, by using the advanced null coordinate $v$ defined by
\be\label{v}
dv =cdT+\left(1-\frac{2m}{r}-\frac{\Lambda}{3} r^2\right)^{-1} dr
\ee 
the (extended) metric becomes
\be\label{kot2}
ds^2 = -\left(1-\frac{2m}{r}-\frac{\Lambda}{3} r^2\right) dv^2 +2dv dr +r^2 d\Omega^2 
\ee
with $r\in (0,\infty)$ (alternatively $r\in (-\infty,0)$). The hypersurface $r=r_+$ represents a cosmological horizon, and that with $r=r_-$ a black hole horizon, both of them null hypersurfaces foliated by marginally trapped round spheres. One has
$$
0< r_-< \frac{1}{\sqrt{\Lambda}} < r_+ < \sqrt{\frac{3}{\Lambda}}
$$
the static region given by $r_- < r < r_+$, while the round spheres with $r<r_-$ and those with $r>r_+$ are trapped --see \cite{S,SMilestone} for the terminology and \cite{GH,GP} for further details. Thus, $r$ is a time coordinate in those two regions. The spacetime contains a curvature singularity at $r=0$ unless $m=0$. The conformal diagram of this spacetime is shown in Figures \ref{fig:kot}, \ref{fig:kot1}  and \ref{fig:dS} \cite{LR,GH,GP}.

\begin{figure}[h!]
\includegraphics[width=10cm]{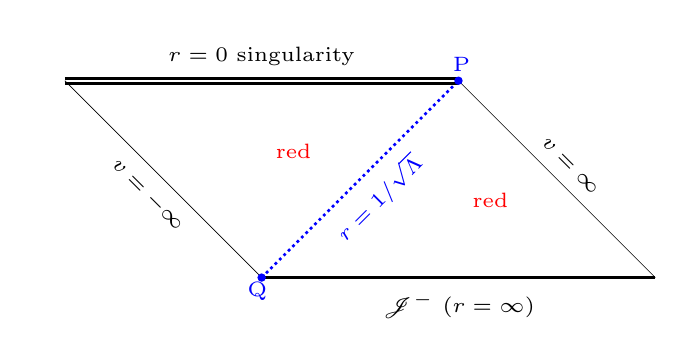}
\caption{{\footnotesize Conformal diagram of the Kottler metric with $9m^2\Lambda=1$ for the portion covered by the coordinates of \eqref{kot2}, same conventions as before. All round spheres are future trapped except those on the blue dotted line, which is a degenerate Killing horizon foliated by marginally trapped spheres. Thus, not to over-red the picture, I have signaled the `red' regions with the word `red'. There is a curvature singularity in the future and past null infinity $\scri^-$. The `points' $P$ and $Q$ marked in the diagram are actually special regions at infinity, so that the null generators of the degenerate horizon are past and future complete. The metric is analytically extendible towards the left and the right by just adding copies of the same diagram, leading to many, or infinite, BH regions. (Again, there is the alternative of making identifications). The topology of $\scri^-$ is $\mathbb{R}\times \mathbb{S}^2$. Observers starting from $\scri^-$ can either end up at the singularity or try to avoid this by reaching $P$ with $r\longrightarrow 3m=1/\sqrt{\Lambda}$ as proper time goes to infinity. However, very few observers in free fall (geodesics) reach $P$ \cite{P}.}}
\label{fig:kotext}
\end{figure}

The limiting possibility is when
$$
\Lambda = \frac{1}{9m^2}
$$
in which case there is only one (double) positive zero of the function $1-\frac{2m}{r}-\frac{r^2}{27m^2} $ given by
$$
r = \frac{1}{\sqrt{\Lambda}} = 3m .
$$
Now, the round spheres with constant $v$ and $r$ are always untrapped except for those with $r = \frac{1}{\sqrt{\Lambda}} = 3m $ which are marginally trapped. The hypersurface $r= \frac{1}{\sqrt{\Lambda}} = 3m $ is a degenerate Killing horizon, and infinity is only reachable for a tiny set of privileged observers \cite{P} ---a subset of the causal geodesics with $T=$const.\ in the original coordinates \eqref{kot} plus the lightlike geodesics on the horizon with $r=3m$. The global structure and general properties of this case were thoroughly analyzed in \cite{P}. The conformal diagram is shown in Figure \ref{fig:kotext}. Even though since the work in \cite{GPe} this extremal BH has been considered many times to actually possess a finite 4-volume at the degenerate horizon, this is a misleading conclusion and the correct interpretation can be found in \cite{Sto}.

\begin{figure}[h!]
\includegraphics[width=10cm]{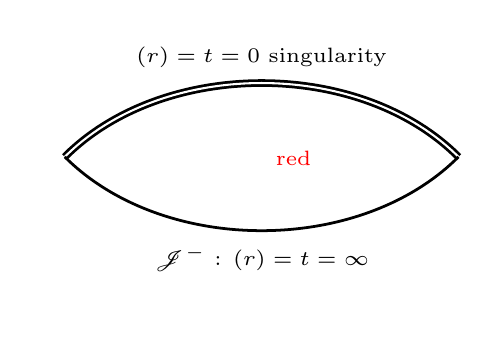}
\caption{{\footnotesize Conformal diagram of the Kottler metric with $9m^2\Lambda>1$ same conventions as before. In this case the original coordinates cover the entire spacetime. It is more visual to call $r \rightarrow t$, as in \eqref{kot3}, because now $r$ is a time coordinate everywhere. In the diagram $t$ decreases towards the future. All round spheres are now future trapped and there are no horizons whatsoever, so the entire diagram is now `red'. The spacetime represents a collapsing universe, with space topology $\mathbb{R}\times \mathbb{S}^2$. There is a curvature singularity in the future and past null infinity $\scri^-$. There is no escape of the singularity, all observers will eventually end up there, so that there is no future null infinity nor asymptotic regions of any kind to the future. The metric is inextendible. The topology of $\scri^-$ is $\mathbb{R}\times \mathbb{S}^2$. The `points' where the singularity and $\scri^-$ touch are not part of the diagram.}}
\label{fig:kot2}
\end{figure}

Finally, there is the case with
$$
\Lambda > \frac{1}{9m^2}
$$
so that in this situation the function $1-\frac{2m}{r}-\frac{\Lambda}{3} r^2$ has no real roots. In this case, there are no horizons and actually the original coordinates of \eqref{kot} cover the entire spacetime. However, as now the coordinate $r$ is a time everywhere, one should better write \eqref{kot} by renaming the coordinates so that is visually clearer:
\be\label{kot3}
ds^2 = -\left(\frac{\Lambda}{3} t^2 +\frac{2m}{t} -1\right)^{-1} dt^2 +\left(\frac{\Lambda}{3} t^2 +\frac{2m}{t} -1\right) dX^2 +t^2d\Omega^2 .
\ee
Notice that $\left(\frac{\Lambda}{3} t^2 +\frac{2m}{t} -1\right)$ is positive everywhere, $t\in(0,\infty)$ and, for compatibility with previous cases, I asusme that $-\partial_t$ is future pointing. Thus, there is a future curvature singularity at $t=0$. The spacetime represents a {\em locally rotationally symmetric} vacuum cosmological contracting model of `Kantowski-Sachs' type \cite{Exact}, included in the general solution first found in \cite{CD}. The metric is inextendible, and the conformal diagram \cite{G} is shown in Figure \ref{fig:kot2}.


\begin{thebibliography}{[10]} 
\bibitem{AMS} Andersson L, Mars M and Simon W, Local existence of dynamical and trapping horizons {\it Phys. Rev. Lett} {\bf 95} (2005) 111102
\bibitem{AMS1} Andersson L, Mars M and Simon W, Stability of marginally outer trapped surfaces and existence of marginally outer trapped tubes, {\it Adv. Theor. Math. Phys.} {\bf 12} (2008) 853 

\bibitem{AMMS} Andersson L, Mars M, Metzger J and Simon W, The time evolution of marginally trapped surfaces {\it Class. Quantum Grav.} {\bf 26} (2009) 085018

\bibitem{AM} Andersson L and Metzger J, The area of horizons and the trapped region, {\it Commun. Math. Phys.} {\bf 290} (2009) 941

\bibitem{ABK} Ashtekar A, Bonga B and Kesavan A, Asymptotics with a positive cosmological constant: I. Basic framework {\it Class. Quantum Grav.} {\bf 32} (2015) 025004
\bibitem{AK} Ashtekar A and Krishnan B, Isolated and dynamical horizons and their applications, {\it  Living Rev. Relativ.} {\bf 7} (2004) 10.
\bibitem{AG} Ashtekar A and Galloway G J, Some uniqueness results for dynamical horizons {\it Adv. Theor. Math. Phys.} {\bf 9} (2005) 1

\bibitem{BE} Bousso R and Engelhardt N, New Area Law in General Relativity, {\it Phys. Rev. Lett.} {\bf 115} (2015) 081301

\bibitem{BE1} Bousso R and Engelhardt N, Proof of a new area law in general relativity, {\it Phys. Rev. Lett.} {\bf 92} (2015) 044031


\bibitem{BeS} Bengtsson I and Senovilla J M M,  Region with trapped surfaces in spherical symmetry, its core, and their boundaries {\it Phys. Rev. D} {\bf 83} (2011) 044012
\bibitem{BJS} Bengtsson I, Jakobson E and Senovilla J M M  Trapped surfaces in Oppenheimer-Snyder black holes {\it Phys. Rev. D} {\bf 88} (2013) 064012
\bibitem{Booth} Booth I, Black hole boundaries {\it Can. J. Phys.} {\bf 83} (2005) 1073--1099

\bibitem{BP} Bi\v{c}\'ak J and Podolsk\'y, Global structure of Robinson-Trautman radiative spacetimes with cosmological constant, {\it Phys. Rev. D} {\bf 55} (1997) 1985

\bibitem{CD} Cahen M and Defrise L, Lorentzian 4-dimensional manifolds with `local isotropy', {\it Commun. Math. Phys.} {\bf 11} (1968) 56.

\bibitem{DJCJ} Deshingar S S, Jhingan S, Chamorro A, and Joshi P S, Gravitational collapse and the cosmological constant, {\it Phys. Rev. D} {\bf 63} (2001) 124005

\bibitem{E} Eichmair M, Existence, regularity, and properties of generalized apparent horizons, {\it Commun. Math. Phys.} {\bf 294} (2010) 745

\bibitem{ESt} Einstein A and Straus E G, The influence of the expansion of space on the gravitation fields surrounding the individual stars {\it Rev. Mod. Phys.} {\bf 17} (1945) 120; erratum {\it ibid.} {\bf 18} (1946) 148
\bibitem{FST} Fayos F, Senovilla J M M and Torres R, General matching of two spherically symmetric spacetimes
{\it Phys. Rev. D} {\bf 54} (1996) 4862.

\bibitem{Frie0} Friedrich H, Existence and structure of past asymptotically simple solutions of Einstein's field equations with positive cosmological constant {\it J. Geom. Phys.} {\bf  3} (1986) 101--117 

\bibitem{GV} Garfinkle D and Vuille C, Gravitational collapse with a cosmological constant , {\it Gen. Rel. Grav.} {\bf 23} (1991) 471-475


\bibitem{G} Geyer K H, Geometrie der Raum-Zeit der Massbestimmung von Kottler, Weyl und Trefftz, {\it Astron. Nachr.} {\bf 301} (1980) 135.

\bibitem{GH} Gibbons G W and Hawking S W, Cosmological event horizons, thermodynamics, and particle creation, {\it Phys. Rev. D} {\bf 15} (1977) 2738

\bibitem{GPe} Ginsbarg P H and Perry M J, Semiclassical perdurance of de Sitter space, {\it Nucl. Phys. B} {\bf 222} (1983) 245

\bibitem{GP} Griffiths J B and Podolsk\'{y} J, {\it  Exact Spacetimes in Einstein's General Relativity} Cambridge Univ. Press. 2009

\bibitem{Haj} Hajicek P, Origin of Hawking radiation
{\it Phys. Rev. D} {\bf 36} (1987) 1065


\bibitem{Hay} Hayward S A, General laws of black-hole dynamics {\it Phys. Rev. D}, {\bf 49} (1994) 6467 

\bibitem{HSN} Hayward S A, Shiromizu T and Nakao K, A cosmological constant limits the size of black holes {\it Phys. Rev. D} {\bf 49} (1994) 5080-85

\bibitem{H} Huang W-H, Hawking radiation of a quantum black hole in an inflationary universe {\it Class. Quantum Grav} {\bf 9} (1992) 1199

\bibitem{Kod} Kodama H, Conserved Energy Flux for the Spherically Symmetric System and the Backreaction Problem in the Black Hole Evaporation, {\it Progress of Theoretical Physics} {\bf 63} (1980)  1217

\bibitem{Kot} Kottler, F, \"Uber die physikalischen Grundlagen der Einsteinschen Gravitationstheorie, {\it Ann. Phys.} (Berlin) {\bf 56} (1918) 401--461

\bibitem{K} Kuroda Y, Naked singularities in the Vaidya spacetime {\it Prog. Theor. Phys.} {\bf 72} (1984) 63

\bibitem{L} Lake K, Gravitational collapse of dust with a cosmological constant, {\it Phys. Rev. D} {\bf 62} (2000) 027301

\bibitem{LR} Lake K and Roeder R C, Effects of a nonvanishing cosmological constant on the spherically symmetric vacuum manifold {\it Phys. Rev. D} {\bf 15} (1977) 3513

%
%
\bibitem{Mallet} Mallet R L, Radiating Vaidya metric imbedded in de Sitter space, {\it Phys. Rev. D} {\bf 31} (1985) 416

\bibitem{Mallet1} Mallet R L, Evolution of evaporating black holes in an inflationary universe {\it Phys. Rev. D} {\bf 33} (1986) 2201; erratum, {\it ibid.} {\bf 34} (1986) 664

\bibitem{MS} Markovic D and Shapiro S L, Gravitational collapse with a cosmological constant {\it Phys. Rev. D} {\bf 61} (2000) 084029 

\bibitem{MMV} Mars M, Mena F C, and Vera R., Review on exact and perturbative deformations
of the Einstein-Straus model: uniqueness and rigidity results {\it Gen. Relativ. Gravit.} {\bf 45} (2013) 2143-2173

\bibitem{MaSe} Mars M and Senovilla J M M , Trapped surfaces and symmetries, {\it Class. Quantum Grav.} {\bf 20} (2003) L293

\bibitem{M} Matyjasek J, Cosmological constant and black hole evaporation {\it Phys. Lett. A} {\bf 120} (1987) 179



\bibitem{N} Nakao K, The Oppenheimer-Snyder spacetime with a cosmological constant, {\it Gen. Rel. Grav.} {\bf 24} (1992) 1069

\bibitem{OS} Oppenheimer J R and Snyder H, On continued gravitational 
contraction {\it Phys. Rev.} {\bf 56} (1939)  455-459

\bibitem{Perlmutter1999}
Perlmutter, S.;   Aldering, G.; Goldhaber, G.; Knop, R.A.; Nugent, P.; Castro, P.G.; Deustua, S.; Fabbro, S.; Goobar, A.; Groom, D.E.; et~al., Measurements of $\Omega$ and $\Lambda$ from 42 High-Redshift Supernovae
{\em Astrophys. J.} {\bf 517} (1999) 565--586, https://doi.org/10.1086/307221.

\bibitem{Planck} Planck Collaboration, Planck 2018 results VI. Cosmological parameters, {\it Astronomy and Astrophysics}, {\bf 641} (2020) A6 

\bibitem{P} Podolsk\'{y} J, The structure of the extreme Schwarzschild-de Sitter spacetime, {\it Gen. Rel. Grav.} {\bf 31} (1999) 1703.

\bibitem{PS} Podolsk\'y J and Sv\'\i tek O, Radiative spacetimes approaching the Vaidya metric, {\it Phys. Rev. D} {\bf 71} (2005) 124001

\bibitem{Riess1998}
Riess, A.G.;  Filippenko, A.V.; Challis, P.; Clocchiatti, A.; Diercks, A.; Garnavich, P.M.; Gilliland, R.L.; Hogan, C.J.; Jha, S.; Kirshner, R.P.; et~al., Observational Evidence from Supernovae for an Accelerating Universe and a Cosmological Constant,
{\em Astron. J.}  {\bf 116}, (1998) 1009--1038, https://doi.org/10.1086/300499.

\bibitem{SA} Sharif M, and Ahmad Z, Gravitational perfect fluid collapse with cosmological constant {\it Mod. Phys. Lett. A} {\bf 22} (2007) 1493--1502

\bibitem{SA1} Sharif M and Ahmad Z, Gravitational dust collapse with cosmological constant, {\it World Applied Sciences Journal} {\bf  16} (2012) 1516-1520

\bibitem{S} Senovilla J M M, Trapped surfaces. {\it Int. J. Mod. Phys. D} {\bf  20} (2011) 2139– 2168

\bibitem{SMilestone} Senovilla J M M and Garfinkle D, The 1965 Penrose singularity theorem, {\it Class. Quantum Grav.} {\bf 32} (2015) 124008

\bibitem{ST} Senovilla J M M and Torres R, Particle production from marginally trapped surfaces of general spacetimes, {\it Classical and Quantum Gravity} {\bf 32} (2015) 085004; Corrigendum: {\it Class. Quantum Grav.} {\bf 32} (2015) 189501

\bibitem{Simon} Simon W, Bounds on area and charge for marginally trapped surfaces with cosmological constant {\it Class. Quantum Grav} {\bf 29} (2012) 062001

\bibitem{Exact} H. Stephani, D. Kramer, M.~A.~H. MacCallum, C. Hoenselaers and E. Herlt. \emph{Exact Solutions of Einstein's Field Equations}, 2nd ed. Cambridge Monographs on Mathematical Physics, Cambridge University Press, 2003.

\bibitem{Sto} Stotyn S, A tale of two horizons, {\it Can J. Phys.} {\bf 93} (2015) 995

\bibitem{VS} Vaidya P C and Shah K B, A radiating mass particle in an expanding universe {\it Proc. Nat. Inst. of Sci. (India)} {\bf 23} (1957) 534

\bibitem{WM} Wagh S M and Maharaj S, Naked Singularity of the Vaidya-de Sitter Spacetime and Cosmic Censorship Conjecture, {\it Gen. Rel. Grav.} {\bf 31} (1999) 975

\bibitem{Wald} Wald, R M, {\it General Relativity}. The University of Chicago Press, 1984.
\bibitem{Wald1} Wald R M, Gravitational collapse and cosmic censorship, in {\it Black holes, gravitational radiation and the Universe}, B R Iyer and B Bhawal eds. (Springer, Berlin) 1998

\bibitem{W} Woolgar E, Bounded area theorems for higher-genus black holes {\it Class. Quantum Grav.} {\bf 16} (1999) 3005

\bibitem{ZYR} Zhao Zheng, Yang Cheng-quan, and Ren Qin-an, Hawking Effect in Vaidya-de Sitter Space-time {\it Gen. Rel. Grav} {\bf 26} (1994) 1055

\end{thebibliography}
\end{document}